\documentclass[11pt, a4paper]{article} 

\usepackage[ansinew]{inputenc}
\usepackage{amsmath, amssymb, graphics, amsthm}
\usepackage{epsfig}
\usepackage{color, xcolor}
\usepackage{fancyhdr} 
\usepackage{manfnt}
\usepackage[T1]{fontenc}

\usepackage[normalem]{ulem}

\newcommand{\ket}[1]{\left| #1 \right\rangle}

\newcommand{\braket}[2]{\left\langle \vphantom {#1 #2} #1 \hphantom{|} \right| \left. \vphantom {#1 #2} #2 \right\rangle}

\makeatletter
\@addtoreset{equation}{section}
\makeatother

\newcommand{\be}{\begin{equation}}
\newcommand{\ee}{\end{equation}}
\newcommand{\ba}{\begin{eqnarray}}
\newcommand{\ea}{\end{eqnarray}}

\def\pb#1{\rlap{\lower1.5ex\hbox{$\longleftarrow$}}{#1}}
\def\dpb#1{\rlap{\lower1.5ex\hbox{$\Longleftarrow$}}{#1}}
\def\spb#1{\rlap{\lower1.0ex\hbox{$\leftarrow$}}{#1}}
\def\sdpb#1{\rlap{\lower1.0ex\hbox{$\Leftarrow$}}{#1}}

\DeclareMathOperator{\sign}{sign}

\title{{\sf An embedding of loop quantum cosmology in $(b, v)$ variables into a full theory context}}
\author{
{\sf N. Bodendorfer}\thanks{{\sf 
norbert.bodendorfer@fuw.edu.pl}}\\
\\
{\sf  Faculty of Physics, University of Warsaw, Pasteura 5, 02-093, Warsaw, Poland}\\
}
\date{{\small\sf May 17, 2016}}

\begin{document} 

\maketitle

{\sf

\begin{abstract}

Loop quantum cosmology in $(b,v)$ variables, which is governed by a unit step size difference equation, is embedded into a full theory context based on similar variables. 
A full theory context here means a theory of quantum gravity arrived at using the quantisation techniques used in loop quantum gravity, however based on a different choice of elementary variables and classical gauge fixing suggested by loop quantum cosmology.
From the full theory perspective, the symmetry reduction is characterised by the vanishing of certain phase space functions which are implemented as operator equations in the quantum theory. The loop quantum cosmology dynamics arise as the action of the full theory Hamiltonian on maximally coarse states in the kernel of the reduction constraints. An application of this reduction procedure to spherical symmetry is also sketched, with similar results, but only one canonical pair in $(b,v)$ form. 

\end{abstract}

}

\section{Introduction}

The issue of performing a symmetry reduction in loop quantum gravity directly at the quantum level has received much attention in recent years \cite{BojowaldSymmetryReductionFor, BojowaldSphericallySymmetricQuantum, EngleRelatingLoopQuantum, BrunnemannSymmetryReductionOf, BianchiTowardsSpinfoamCosmology, AlesciANewPerspective, EngleEmbeddingLoopQuantum, GielenCosmologyFromGroup, HanuschInvariantConnectionsIn}. Given the success in applying loop quantum gravity methods in mini- \cite{BojowaldAbsenceOfSingularity, AshtekarMathematicalStructureOf, AshtekarQuantumNatureOf} and midi-superspaces \cite{KastrupCanonicalQuantizationOf, BojowaldLemaitreTolmanBondi, GambiniLoopQuantizationOf}, this line of work has the important goal of transferring these results to a full theory setting, or to correct them accordingly. Within the canonical theory, an approach based on choosing adapted quantisation variables has allowed to make additional progress along this route more recently: it was shown how a symmetry reduction at the quantum level could be performed for Bianchi I models \cite{BIII} and for spherical symmetry \cite{BLSI}. For Bianchi I models, it could furthermore be shown that the mini-superspace dynamics \cite{AshtekarLoopQuantumCosmologyBianchi} derived in the context of loop quantum cosmology can be reproduced by an appropriate quantisation. In the case of spherical symmetry, this has not been possible so far in full generality due to a more involved structure of the reduced quantum theory. Still, some progress has been made in \cite{BZI}, where the qualitative structure of the mini-superspace quantum algebra could be reproduced from the full theory. 

In this paper, we are going to propose an alternative quantum reduction to both spatially flat isotropic and homogeneous cosmology as well as spherical symmetry by choosing a different set of quantisation variables than in \cite{BIII, BLSI}. 
We achieve two new results: 

1) We embed loop quantum cosmology in $(b,v)$ variables into a full theory context. This is of interest because an embedding in terms of $(c,p)$ variables, as done in \cite{BIII}, requires a non-trivial adaption of the polymerisation scale in the Hamiltonian on the spatial geometry in order to implement the $\bar \mu$ scheme \cite{AshtekarQuantumNatureOf}. 
A formulation of the full theory in $(b,v)$-type variables on the other hand allows one to use maximally coarse quantum states and at the same time have maximally simple and physically viable quantum dynamics in terms of a unit step size difference equation.

2) The methods developed in the case of cosmology can be directly applied to spherical symmetry. Also here, one can reproduce the algebraic relations one obtains from a quantisation of the classically reduced theory from a full theory quantisation. Together with a map of reduced and full theory quantum states, one then obtains the classically reduced quantisation by expressing the classically reduced Hamiltonian as a full theory operator acting on the quantum symmetry reduced full theory states. Due to the more complicated choice of variables in \cite{BLSI}, this result could so far be achieved only qualitatively in the case of spherical symmetry \cite{BZI}.

To describe our results, we choose the formulation to ``embed the classically reduced quantisation into a full theory {\it context}''. By this, we mean that the corresponding full theory has so far not been shown to possess all necessary properties, such as an anomaly-free Hamiltonian constraint with a kernel admitting sufficiently many semiclassically well behaved states. What has been shown is that:
a) The full theory quantum kinematics used here can be obtained from a loop quantum gravity type\footnote{By this we mean a canonical quantisation based on variables such that the smearing dimensions of conjugate pairs add up to the spatial dimension, and that one of the variables in the conjugate pair exists only in exponentiated form, being interpreted as a group element in a certain representation. The quantisation then mimics the techniques used in standard loop quantum gravity, e.g. in the construction of the Hilbert space or the definition of the dynamics. In the current paper, the case of scalar fields treated in \cite{ThiemannQSD5, ThiemannKinematicalHilbertSpaces, AshtekarPolymerAndFock} is mainly relevant.} quantisation of the classical full theory phase space in certain gauges.
b) A suitable set of reduction constraints can be implemented as operators on the Hilbert space. 
c) A preferred set of states in the kernel of the reduction constraints is in a one-to-one correspondence with the quantum states obtained from a corresponding quantisation of a classically reduced model.
d) A preferred set of full theory operators corresponding to certain operators in the classically reduced quantisation have the same algebraic properties as the operators obtained from quantising the classically reduced theory when acting on symmetry reduced states. 
e) The Hamiltonian of the classically reduced theory expressed as an operator on the full theory Hilbert space acts in the same way as it would when quantising the classically reduced theory. 
f) There exists a plausible quantisation of the full theory Hamiltonian such that it reduces to the classically reduced one when acting on maximally coarse states. This last step is sketched explicitly only in the case of cosmology in this paper. 

Given these results, we find this choice of words fitting. It was our aim to avoid to talk about an embedding in {\it the} full theory, since such a statement implies that a choice of variables and quantisation map has been made once and for all. Still, it seems reasonable to refer to the full theories relevant in this paper as loop quantum gravity (in the wider context), since the quantisation methods used are the same, while the choice of variables and the corresponding smearings are different. 

It is interesting to ask whether the theory constructed this way is equivalent to standard loop quantum gravity. The typical answer to this question would be that different quantisations usually differ in the details, so that one would expect an answer in the negative. However, it turns out to be very hard to make this question precise, as one does not compare two quantisations of the same algebra, but two quantisations of two different algebras of elementary phase space functions, with one of them referring classically to a certain choice of coordinates via the gauge fixing. In particular, there does not seem to be a canonical, i.e. regularisation-independent way in constructing a map from operators in one quantum theory to the other. 
What one could do is to just compare the theories for a certain subset of operators where one has a plausible regularisation in one theory, whereas the operator is among the elementary functions in the second theory. In the present case, an example of this is the volume of the whole universe, which also does not depend on the choice of spatial coordinates. Here, one finds differences in the spectrum (a countable, possibly dense subset of $\mathbb R^+_0$ in standard LQG vs. $\mathbb R^+_0$ in our case), which gives a hint that the two theories are in fact different mathematically, but not necessarily given a finite measurement uncertainty. See also the second remark in section \ref{sec:Comments}.

This paper is organised as follows:\\
We consider the case of a reduction to spatially flat homogenous and isotropic cosmology in the main part of the paper. In section \ref{sec:Strategy}, we will explain our general strategy for the reduction. The classical preparations necessary will be performed in section \ref{sec:Preparations}. In section \ref{sec:QuantumKinematics}, we will discuss the quantum kinematics. The imposition of the reduction constraints and the quantum dynamics follow in section \ref{sec:ReductionQuantisation}. We comment on our results in section \ref{sec:Comments} and conclude in section \ref{sec:Conclusion}. Appendix \ref{app:FullHamiltonian} sketches the construction of the full theory Hamiltonian constraint.
In appendix \ref{app:SphericalSymmetry}, we discuss the case of spherical symmetry to avoid too much conceptual repetition in the main text.

\section{General strategy} \label{sec:Strategy}

The general strategy of this paper follows that of \cite{BIII, BLSI}. The main steps are:
\begin{enumerate}
	\item {\bf Formulate the classical theory in a gauge adapted to the intended reduction.} \\
		By choosing different gauges, one obtains more freedom in the choice of variables and the imposition of reduction constraints can be significantly easier. In the case of reducing to a cosmological model, the choice of gauge will be the diagonal metric gauge. 
		
	\item {\bf Choose adapted variables.}\\
		Prior to quantisation, we choose variables such that the reduction at the quantum level becomes as easy as possible. In particular, we make a choice such that the $\bar \mu$-scheme of loop quantum cosmology \cite{AshtekarQuantumNatureOf} translates into a unit step size difference equation. We thus choose a full theory analogue of the $(b,v)$-variables \cite{AshtekarQuantumNatureOf}.  

	\item {\bf Identify suitable reduction constraints.}\\
		A set of phase space functions which vanish in the symmetry reduced sector of phase space will be identified. These functions, denoted as {\it reduction constraints}, will later be quantised and their vanishing imposed as operator equations. These constraints need only be necessary, but not sufficient, to achieve the symmetry reduction at the classical level, and they are chosen with hindsight as to simplify the quantum theory as much as possible.

	\item {\bf Quantise the reduced\footnote{``Reduced phase space'' is the standard terminology to refer to the phase space obtained after solving a set of second class constraints, e.g., after gauge fixing. In the present context, this should not be confused with reducing degrees of freedom. This step will be enforced at the quantum level via the imposition of ``reduction constraints''.} phase space.}\\
		We quantise the reduced phase space, i.e. the phase space obtained after gauge fixing, but before imposing any reduction constraints. The resulting Hilbert space can in principle still serve to formulate full quantum general relativity in the diagonal metric gauge.
		
	\item {\bf Impose the reduction constraints.}\\
		The reduction constraints will be quantised. It will turn out that they constitute a second class set. A first class set needs to be picked consistently and imposed as strong operator equations. 
				
	\item {\bf Compare to loop quantum cosmology.}\\
		The dynamics determined by the Hamiltonian constraint will be compared to those of loop quantum cosmology. The interface for the dynamical degrees of freedom in the classically and quantum reduced theories is provided by full theory operators commuting with the reduction constraints. 

\end{enumerate}

\section{Classical preparations} \label{sec:Preparations}

The main classical ingredient in our derivation will be an adapted choice of variables such that the reduction constraints can be implemented as easily as possible at the quantum level. We start with the usual ADM phase space \cite{ArnowittTheDynamicsOf} including a massless scalar field. The non-vanishing Poisson brackets between the spatial metric $q_{ab}$, its conjugate momentum $P^{ab}$, the scalar field $\phi$, and its momentum $P_\phi$ are given by
\be
	\left\{q_{ab}(x), P^{cd}(y) \right\}  =  \delta^{(3)}(x,y) \, \delta_{(a}^c \delta_{b)}^d, ~~~~~~  \left\{\phi(x), P_\phi(y) \right\}  =  \delta^{(3)}(x,y)  \text{.} \label{eq:ADMPoissonBracket}
\ee
The phase space is subject the spatial diffeomorphism constraint 
\be
	C_a = -2 \nabla_b P^{b} {}_a + P_\phi \partial_a \phi \approx 0
\ee
as well as the Hamiltonian constraint
\be
	H =  - \frac{P^2} {3 \sqrt{q}} + \frac{4}{\sqrt{q}} P_{ab}^{\text{tf}} P^{ab}_{\text{tf}} - \frac{1}{2} \sqrt{q} R+ \frac 12 P_\phi^2 / \sqrt{q} + \sqrt{q} q^{ab} \partial_a \phi \partial_b \phi \approx 0  \text{,} \label{eq:FullHamiltonian}
\ee 
where $P = P^{ab} q_{ab}$ and $P^{ab}_\text{tf} = P^{ab} - \frac 13 q^{ab}P$ denote the trace and traceless parts of $P^{ab}$.
We restrict the topology to be a 3-torus in order not to have to worry about boundary terms or infinite volume.
Following \cite{BIII}, we impose the diagonal metric gauge $q_{ab} = \text{diag}(q_{xx}, q_{yy}, q_{zz})$ for the spatial diffeomorphism constraint\footnote{Global issues concerning the accessibility of the gauge are being neglected here, i.e. we restrict to a part of phase space where the gauge is accessible. While diagonalisation of a 3-metric is known to be always possible locally around a given point on the spatial slice \cite{DeTruckExistenceOfElastic}, this restricts the class of spacetimes that can be treated in the current formalism.}. The reduced phase space is then parametrised by the diagonal components of $q_{ab}$ and $P^{ab}$, which inherit the canonical Poisson bracket \eqref{eq:ADMPoissonBracket}. The off-diagonal components of $P^{ab}$ are solved for by $C_a = 0$. A subset of spatial diffeomorphisms preserving the diagonal gauge, which are later merged with the reduction constraints, remain first class. 

We are now in a position to discuss the constraints leading to a reduction to homogeneous and isotropic cosmology with vanishing spatial curvature, i.e. a $k=0$ FRW model. There, in a suitable coordinate system, we have $q_{ab} \propto  \delta_{ab}$ and $P^{ab} \propto \delta^{ab}$. The two proportionality constants are independent of the spatial coordinate and constitute the left over degrees of freedom. We will now derive a set of phase space functions which vanish in this symmetry reduced sector and denote them as {\it reduction constraints} in the following. For this purpose, and also to be able to quantise more easily later on, we will first choose a new set of canonical variables in the gravitational sector inspired by the $(b,v)$-variables used in loop quantum cosmology \cite{AshtekarQuantumNatureOf}.

We define 
\begin{align}
		\alpha &:= \sqrt{q_{xx} q_{yy} q_{zz}} ~ &P_\alpha &:= \frac{2 P^{xx} q_{xx}}{3 \sqrt{q_{xx} q_{yy} q_{zz}}} + \frac{2 P^{yy} q_{yy}}{3 \sqrt{q_{xx} q_{yy} q_{zz}}} + \frac{2 P^{zz} q_{zz}}{3 \sqrt{q_{xx} q_{yy} q_{zz}}} \label{eq:NewVars1}\\
		\beta &:=  P^{xx}q_{xx} - P^{yy} q_{yy}~~ & P_\beta &:= \frac{1}{2} \log \frac{q_{yy}}{q_{xx}}  \label{eq:NewVars2}\\
		\gamma &:= P^{xx}q_{xx} - P^{zz} q_{zz}  & P_\gamma &:= \frac12 \log \frac{q_{zz}}{q_{xx}}  \label{eq:NewVars3}
\end{align} 
and note that the logarithms are well defined since positive definiteness of the metric forces $q_{xx} >0$, $q_{yy} >0$, and $q_{zz} >0$.
As the notation suggests, the non-vanishing Poisson brackets are 
\be
	\left\{  \alpha (x),   P_\alpha(y) \right\} =  \delta^{(3)}(x,y), ~~~   \left\{ \beta (x),   P_\beta(y)   \right\} =   \delta^{(3)}(x,y),   ~~~ \left\{   \gamma (x),   P_\gamma(y)   \right\} =   \delta^{(3)}(x,y)   \text{.}
\ee
We verify immediately that for a homogeneous and isotropic universe $\beta = \gamma = P_\beta = P_\gamma = 0$. $\alpha$ is a measure of the total volume $V = \int_\Sigma d^3x \,\alpha$ of the universe. It is thus clear that $(P_\alpha, \alpha)$ correspond to the $(b,v)$ variables of loop quantum cosmology if we smear $\alpha$ over a three-volume, e.g. the total universe, and construct point holonomies from $P_\alpha$. We could also introduce a Barbero-Immirzi parameter at this point, but will only do so later in equation \eqref{eq:RelBNuAlpha} when making an explicit comparison to loop quantum cosmology.

Independently of these constraints, we can still impose $P^{a\neq b}=0$, i.e. demanding that the off-diagonal components of $P^{ab}$ vanish.
It was shown in \cite{BIII} that imposing the constraint $P^{a\neq b}[\mu_{a \neq b}]=0$, i.e. $P^{a \neq b}$ smeared against arbitrary smearing functions $\mu_{a \neq b}$ vanishes, translates to the condition 
\begin{align}
	0=\left. P^{ab} \mathcal L_{N} q_{ab} + P_\phi \mathcal L_{N} \phi \right|_{q_{a \neq b} = 0, ~ P^{a \neq b} = 0} =~& P^{xx} N^c \partial_c q_{xx} + P^{yy} N^c \partial_c q_{yy} + P^{zz} N^c \partial_c q_{zz} \label{eq:PanotbReduction} \\
		&+ 2 P^{xx} q_{xx} \partial_x N^x + 2 P^{yy} q_{yy} \partial_y N^y+2 P^{zz} q_{zz} \partial_z N^z \nonumber  \\
		&+ P_\phi N^c \partial_c \phi \nonumber
\end{align}
for vector fields $N^a$ determined by $\mu_{a \neq b}$. Sampling over all $\mu_{a \neq b}$ then samples over all\footnote{We restrict to a part of phase space where the construction of $N^a$ as given in \cite{BIII} works globally. While this might exclude a measure 0 from the vector fields produced via $P^{a \neq b}=0$, we will still simply implement the reduction as all vector fields and consider $P^{a \neq b}=0$ as a subset thereof. The diffeomorphisms preserving the gauge conditions are also included in this set. Footnote \ref{ftn:SuperposeConstraints} comments on a related point. As a logical cross-check, we can classically truncate the ADM formulation by setting $q_{a \neq b} = P^{a \neq b} = 0$, arriving at the same system.} $N^a$. The action of \eqref{eq:PanotbReduction} is somewhat complicated since the individual variables are transforming with different density weights for different components of $N^a$. However, we can express \eqref{eq:PanotbReduction} in terms of our new variables and modify it by dropping all terms proportional to $\beta, \gamma, P_\beta$, and $P_\gamma$, which corresponds to taking superpositions of reduction constraints. We find
\be
	0 = \left. P^{ab} \mathcal L_{N} q_{ab}+P_\phi \mathcal L_{N} \phi \right|_{q_{a \neq b} = 0, ~ P^{a \neq b} = 0, ~ \beta = \gamma = P_\beta = P_\gamma = 0} = P_\alpha \mathcal L_N \alpha + P_\phi \mathcal L_N \phi   \text{,} \label{eq:DiffPalpha}
\ee
where the Lie derivative acts on $\alpha$ as a density and on $\phi$ as a scalar. \eqref{eq:DiffPalpha} is now suitable to be used as a reduction constraint, since it corresponds to implementing spatial diffeomorphisms acting on the canonical pairs $(\alpha, P_\alpha)$ and $(\phi, P_\phi)$.

We conclude that we have found that in a $k=0$ FRW universe, we can impose the conditions
\begin{align}
	&0= \beta = \gamma = P_\beta = P_\gamma, ~~~ \text{as well as}  ~~~ 0=   P_\alpha \mathcal L_N  \alpha + P_\phi \mathcal L_{N} \phi \text{.}
\end{align}
While the first set of conditions follow directly from the form of the metric and its conjugate momentum, the last condition is a consequence of the condition $P^{a \neq b} = 0$, expressed on the reduced phase space via the spatial diffeomorphism constraint. 

It is important to note that these conditions are not sufficient to reduce to homogeneity and isotropy at the classical level, however clearly necessary in the chosen gauge. The choice presented here is made with hindsight as to simplify the quantum theory as much as possible and to be able to perform explicit calculations there, as well as to be able to have a link to full theory quantum dynamics, as discussed in appendix \ref{app:FullHamiltonian}. We comment more on this point in section \ref{sec:Comments}, where also the implementation of a first class subset of sufficient constraints (taking in addition $P_\alpha$ and $\phi$ to be constant) is discussed. Let us also mention that in order to obtain a first class subset, we can simply drop $P_\beta = 0$ and $P_\gamma = 0$ from our list of reduction constraints.

\section{Quantum kinematics} \label{sec:QuantumKinematics}

We will choose the same Hilbert spaces for all four canonical pairs $(\alpha, P_\alpha)$, $(\beta, P_\beta)$, $(\gamma, P_\gamma)$, and $(\phi, P_\phi)$, following \cite{ThiemannQSD5, ThiemannKinematicalHilbertSpaces, AshtekarPolymerAndFock}. 
Exemplarily for $(\alpha, P_\alpha)$, we construct point holonomies from $P_\alpha(x)$, i.e. $h_x^\rho := e^{-i \rho P_\alpha(x)}$ for $\rho \in \mathbb R$. Point holonomies are then the direct analogue of holonomies of the Ashtekar-Barbero connection \cite{AshtekarNewVariablesFor, BarberoRealAshtekarVariables} in standard loop quantum gravity. 
A cylindrical function $f$, the basic element of our Hilbert space, is defined as a function depending on finitely many point holonomies. The scalar product on the kinematical Hilbert space between two cylindrical functions $f$ and $g$ is given by
\begin{align}
	\braket{f}{g}_\text{kin} &= \int \prod_{x \in\Sigma} d \mu_{\text{Bohr}}(x) \bar{f} g \text{.}
\end{align}
The construction proceeds in a similar way for the other three sets of canonical variables. 

While there is no operator directly corresponding to $P_\alpha(x)$, the point holonomies $e^{-i \rho P_\alpha(x)}$ act by multiplication. An operator corresponding to $\alpha$ can be constructed by smearing it over a three-dimensional region $R$, denoted by $\alpha(R) := \int_R d^3x \,\alpha$. Following the standard regularisation procedure, we find
\be
	\widehat{\alpha(R)} \ket{f} = i \int_R \frac{\delta}{\delta P_\alpha(x)} \ket{f} = i \sum_{x \in \Sigma} s(R,x)  \frac{\partial}{\partial P_\alpha(x)} \ket{f} \text{.}
\ee
where
\be
	 s(R,x) := \sign (R) ~ \times ~ \begin{cases} 
												1 &\mbox{if } x \in R \backslash \partial R \\ 
												1/2 & \mbox{if } x \in \partial R \\  
												0 & \mbox{otherwise,}					
											\end{cases} 
\ee
and $\sign(R)$ denotes the orientation of $R$ in $\Sigma$.
As an example, in the simple case $f = h_x^\rho = e^{-i \rho P_\alpha(x)}$ with $x$ in the interior of $R$ and $\sign(R)=1$, we get 
\be
	\widehat {\alpha(R)} \ket {h_x^\rho} = \rho \ket {h_x^\rho} \text{.}
\ee
Again, also this construction can be repeated verbatim for the other three canonical pairs. 

At this point, we could discuss how to implement the full theory Hamiltonian constraint on the Hilbert space. Since just a small part of this operator will survive the reduction procedure, we will  comment on the full operator only briefly in appendix \ref{app:FullHamiltonian}. The part of the full Hamiltonian relevant for the reduced sector will be discussed in the next section.

\section{Reduction constraints and quantum dynamics} \label{sec:ReductionQuantisation}

\subsection{Symmetry reduced quantum states} \label{sec:ReductionConstraintsOperators}

It will be our strategy to implement a first class subset of all reduction constraints as operators on the Hilbert space. The remaining reduction constraints, second class partners to the first class subset, will be dropped\footnote{\label{ftn:GU}According to the procedure of gauge unfixing \cite{MitraGaugeInvariantReformulation, VytheeswaranGaugeUnfixingIn}, we then would have to modify all remaining operators of the theory, including the Hamiltonian constraint, by adding terms proportional to the second class partners in such a way that the operators commute with the first class subset. In our case, this would for example mean to drop all dependence on $P_\beta$ and $P_\gamma$ from the Hamiltonian. Since those terms will drop anyway when operating on the symmetry reduced quantum states, we will not explicitly perform this procedure here.}.

We first choose to include $\widehat {\beta(R)}$ and $\widehat {\gamma(R)}$ for arbitrary regions $R$ in our first class subset. Accordingly, polymerised approximates for $P_\beta$ and $P_\gamma$ cannot be contained in the first class subset. A vanishing action of $\widehat {\beta(R)}$ and $\widehat {\gamma(R)}$ on cylindrical functions simply means that they do not depend on point holonomies of $P_\beta$ and $P_\gamma$. 

Next, we implement the spatial diffeomorphisms generated by \eqref{eq:DiffPalpha}. 
This can be done in full analogy to \cite{AshtekarQuantizationOfDiffeomorphism} by a rigging map construction, see also \cite{ThiemannKinematicalHilbertSpaces, SahlmannExploringTheDiffeomorphism}. 
The main idea is to implement all finite diffeomorphisms instead of their generator, since the action of finite diffeomorphisms is well understood on the kinematical Hilbert space. 
We will not spell out the details here and just remark that, morally, a rigging map construction can be performed which effectively diffeomorphism-averages the point holonomies contained in a cylindrical function $f$ over the spatial slice. Diffeomorphism invariant states $\eta[f]$ are then elements of the dual of the Hilbert space. The simplest possible example is to consider the cylindrical function $f = h^\rho_x = e^{-i \rho P_\alpha(x)}$ for some $x\in\Sigma$. $\eta[f]$ turns out to be the element of the dual of the Hilbert space for which
\be
	\eta[h^\rho_x] (h^{\rho'}_y) = \sum_{z \in \Sigma}\braket{h^\rho_z}{h^{\rho'}_y}  =  \delta_{\rho, \rho'} ~ \forall ~ y \in \Sigma \text{.}
\ee
Clearly, $\eta[h^\rho_x]$ is diffeomorphism invariant.
The new scalar product on the diffeomorphism invariant Hilbert space is given by 
\begin{align}
	\braket{\eta[h^\rho_x]}{\eta[h^{\rho'}_y]}_\text{diff} &= \delta_{\rho,\rho'} \text{.}
\end{align} 
It turns out that this simplest example is already sufficient to establish the relation to loop quantum cosmology. We will therefore not discuss more elaborate examples here and refer to the above type of state as a {\it single vertex state}. The implementation of the reduction constraint \eqref{eq:DiffPalpha} thus leads to the above type of diffeomorphism invariant distributions for both the $(\alpha, P_\alpha)$ and $(\phi, P_\phi)$ sectors, i.e. the averaging is performed based on a state with point holonomies  $h^{\rho_\alpha}_x$ for the $(\alpha, P_\alpha)$ and $h^{{\rho_\phi}}_x$ for the $(\phi, P_\phi)$ sectors at coinciding points. 
Since $\phi$ transforms as a scalar under \eqref{eq:DiffPalpha}, the point holonomies in the $(\phi, P_\phi)$ sector are given by $h^{\rho_\phi}_x := e^{i \rho_\phi \phi(x)}$, with the different sign in the exponent chosen so that $\hat P_\phi$ multiplies by $\rho_\phi$.
For brevity, we will denote the resulting state as $h^{\rho_\alpha, \rho_\phi}_{\text{diff}}$, with 
\be
	\braket{h^{\rho_\alpha, \rho_\phi}_{\text{diff}}}{h^{\rho_\alpha', \rho_\phi'}_{\text{diff}}}_\text{diff} = \delta_{\rho_\alpha,\rho'_\alpha} \delta_{\rho_\phi,\rho'_\phi} \text{.} \label{eq:ScalarProductRhoPhi}
\ee

Let us now turn to diffeomorphism invariant operators. As a first example, $\widehat{\alpha(\Sigma)}$ is a diffeomorphism invariant operator because $\Sigma$ is mapped to $\Sigma$ by any diffeomorphism. Its action is therefore well defined on the diffeomorphism invariant states, in particular we have 
\be
	\widehat{\alpha(\Sigma)} \ket{h^{\rho_\alpha, \rho_\phi}_{\text{diff}}} = \rho_{\alpha}  \ket{h^{\rho_\alpha, \rho_\phi}_{\text{diff}}} \text{.} \label{eq:OperatorAlpha}
\ee
Later in this paper, we will change \eqref{eq:OperatorAlpha} to correspond to the absolute value of $\alpha$ to mimic the derivation in loop quantum cosmology and to ensure that $\alpha(\Sigma)$ becomes a positive operator. 

Similarly, we have 
\be
	\widehat{P_\phi(\Sigma)} \ket{h^{\rho_\alpha, \rho_\phi}_{\text{diff}}} = \rho_{\phi}  \ket{h^{\rho_\alpha, \rho_\phi}_{\text{diff}}} \text{.} \label{eq:QOPphi}
\ee
Next, we need to construct a diffeomorphism invariant operator containing $P_\alpha$. In order to do this, we integrate $\frac 1\lambda \sin(\lambda P_\alpha ) \alpha$ over all of $\Sigma$, and choose a factor ordering where $\alpha$ is on the right. $\lambda \in \mathbb R^+$ sets the polymerisation scale, which we need to introduce because $P_\alpha$ has to be approximated by point holonomies\footnote{Approximation of connection-type variables via holonomies is a standard ingredient of loop quantum gravity, necessitated by the non-existence of an operator corresponding to the connection, whereas holonomies do exist on the Hilbert space. The underlying reason is the weak discontinuity of the scalar product, in this case for limits of $\rho$.}. 
It is then straightforward to promote $(\frac 1\lambda \sin(\lambda P_\alpha) \alpha)(\Sigma)$ to a diffeomorphism invariant operator as
\be
	\widehat{\frac 1\lambda \left(\sin(\lambda P_\alpha ) \alpha \right)(\Sigma)} \ket{h^{\rho_\alpha, \rho_\phi}_{\text{diff}}} = \frac{\rho_{\alpha}}{2 i \lambda} \left( \ket{h^{\rho_\alpha-\lambda, \rho_\phi}_{\text{diff}}}- \ket{h^{\rho_\alpha+\lambda, \rho_\phi}_{\text{diff}}} \right) \text{.}  \label{eq:QOPalphaalpha}
\ee
The reason for choosing to order $\alpha$ to the right becomes clear when looking at the details of the regularisation: the action on any state is non-trivial only at a point where the dependence on $e^{-i P_\alpha}$ is non-trivial. 
If we would have ordered $\sin (\lambda P_\alpha)$ to the right, then the operator would have a non-trivial action at any point in $\Sigma$ and not preserve the set of single vertex states. This feature is similar to the property of the Hamiltonian constraint acting only on non-trivial vertices \cite{ThiemannQSD1}. 
We could now also define an operator corresponding to $\frac 1\lambda \sin (\lambda \phi) \alpha$ in a similar way. Since this operator will not be needed for the final Hamiltonian, we will refrain from doing so. 

While the implementation of diffeomorphism invariance here is technically the same as in standard LQG, it is important to stress again that our diffeomorphisms arise from imposing reduction constraints, as opposed to standard LQG, where they are removing gauge degrees of freedom.

\subsection{Quantum dynamics} \label{eq:QuantisationDynamics}

We are now in a position to discuss the quantum dynamics. For this, we need to construct a Hamiltonian constraint, or true Hamiltonian operator if we choose to deparametrise classically. In this section, we will take a shortcut and simply take the classically reduced minisuperspace Hamiltonian and implement it as an operator on the full Hilbert space in a form which allows for deparametrisation both at the classical and quantum level (one simply follows the same route that one would choose in LQC). In general however, it would be more satisfactory to first construct the full Hamiltonian constraint without considering the reduction constraints, implement it on the full Hilbert space, and then show that it preserves the reduced quantum states. More discussion on this issue, as well as an example of such a construction based on a fixed lattice regularisation, is given in appendix \ref{app:FullHamiltonian}.

The classically reduced Hamiltonian can be easily obtained form \eqref{eq:FullHamiltonian} by setting $0= \beta = \gamma = P_\beta = P_\gamma = P^{a \neq b}$ and dropping the three-dimensional Ricci scalar $R$, as well as all spatial derivatives. Using $P = \frac 32 P_\alpha \alpha$ and $\sqrt{q} = \alpha$, we are left with 
\be
	\frac{P_\phi^2 }{2 \alpha} - \frac 34 P_\alpha^2 \alpha \approx 0\text{,} \label{eq:HamREducedNonDepara}
\ee
which can be rewritten as
\be
	P_\phi^2 \approx \frac 32 \left(P_\alpha |\alpha|\right)^2 \text{,} \label{eq:HamiltonianReducedDepara}
\ee
corresponding to the choice of lapse $N = v$. Our choice of lapse will make the quantum theory most simple, $N=1$ as an alternative choice is discussed below.
The minisuperspace Hamiltonian is now obtained by taking square roots, integrating the densities $P_\alpha |\alpha|$ and $P_\phi$ over $\Sigma$, and squaring again: 
\be
	\left( \int_\Sigma d^3x \, P_\phi \right)^2 \approx \frac 32  \left(  \int_\Sigma d^3x \, P_\alpha |\alpha|\right)^2.
\ee
In order to promote this classical constraint equation to an operator on the reduced Hilbert space, we furthermore have to express $P_\alpha$ in terms of point holonomies, i.e. we polymerise it as
\be
	\left( \int_\Sigma d^3x \, P_\phi \right)^2 \approx  \frac{3}{2 \lambda^2}  \left( \int_\Sigma d^3x \, \sin \left( \lambda P_\alpha \right)| \alpha |\right)^2 \label{eq:HamPoly} \text{.}
\ee
It is now straight forward to quantise \eqref{eq:HamPoly} by using the previously derived operators \eqref{eq:QOPphi} and \eqref{eq:QOPalphaalpha} as
\be
	\widehat{ P_\phi(\Sigma)}^2 \ket{h^{\rho_\alpha, \rho_\phi}_{\text{diff}}} = \frac{3}{2\lambda^2} \left(\widehat{ \left(\sin(\lambda P_\alpha ) |\alpha| \right)(\Sigma)}  \right)^2 \ket{h^{\rho_\alpha, \rho_\phi}_{\text{diff}}} \text{.} \label{eq:QuantumHamiltonian}
\ee

Let us now discuss the choice of $\lambda$, which, a priori, can be any function of $\rho_\alpha$. The only physical criterion that we can apply is that the approximation $P_\alpha \approx \sin (\lambda P_\alpha) / \lambda$ has to be good in the regime where we do not expect quantum gravity effects to be present. Quantum gravity effects should become relevant as soon as the matter energy density $\rho
_m$ reaches Planck density. In our case, $\rho_m \sim P^2_\alpha$, so that for $\rho_m \ll 1$, we also have $P_\alpha \ll 1$. Furthermore, we do not expect that the onset of quantum effects depends on the total volume of the universe, but just on the local energy density. Therefore, choosing $\lambda$ to be a fixed real positive number at the order of $1$ is singled out. Within loop quantum cosmology, a heuristic derivation of the precise value of $\lambda$ from the area gap of standard SU$(2)$ loop quantum gravity can be given \cite{AshtekarQuantumNatureOf}, agreeing with our conclusion here. It is also interesting to remark that a different choice of variables, such as usual connections integrated along paths, necessitates a different, state-dependent choice of $\lambda$ \cite{AshtekarLoopQuantumCosmologyBianchi}. A similar argument as the one presented here then also applies \cite{BIII}.

A close inspection shows that equation \eqref{eq:QuantumHamiltonian} is, up to some details to be discussed below, the difference equation that generates the dynamics in loop quantum cosmology formulated using the $\bar \mu$-scheme \cite{AshtekarQuantumNatureOf}. First, we can explicitly change variables in our construction to the $(b,\nu)$-variables\footnote{The difference between $\nu$ \cite{AshtekarRobustnessOfKey} and $v$ \cite{AshtekarQuantumNatureOf} is only a numerical prefactor. While the name $v$ is more suggestive, we choose to use $\nu$ in this section to simplify the comparison to \cite{AshtekarRobustnessOfKey}.} given in \cite{AshtekarRobustnessOfKey} to show this: the definitions (with $\gamma$ here being the Barbero-Immirzi parameter) 
\be 
	\nu :=   \int_\Sigma d^3x \, \tilde  \nu, ~~~~~~ \tilde \nu :=  \frac{4 \alpha}{\gamma}, ~~~~~ b = - \frac{\gamma}{2} P_\alpha \label{eq:RelBNuAlpha}
\ee
yield $\{ b, \nu \} = 2$ and the polymerised minisuperspace Hamiltonian constraint reads
\be
	\left( \int_\Sigma d^3x \, P_\phi \right)^2 \approx \frac {3 \pi G}{ \lambda^2} \left( \int_\Sigma d^3x \, \sin (\lambda b) | \tilde \nu| \right)^2 \text{,} \label{eq:HamiltonianReducedDeparabnu}
\ee
where we restored units using our convention $8 \pi G = 1$. On the states $\ket{\nu, \phi} = e^{i \nu b / 2} f(\phi)$ for some $f$ and after the same quantisation procedure as above and with the same factor ordering, \eqref{eq:HamiltonianReducedDeparabnu} acts as
\begin{align}
	\partial_\phi^2 \ket{\nu, \phi} &= \frac {3 \pi G}{4 \lambda^2} |\nu| \biggl( |\nu+2 \lambda| \ket{\nu+4, \phi} + |\nu-2 \lambda| \ket{\nu-4, \phi} -  (|\nu+2\lambda| + |\nu-2\lambda|) \ket{\nu, \phi} \biggr)\nonumber\\
	& =: - \Theta \ket{\nu, \phi}  \text{,} \label{eq:bnuDifferenceEquation}
\end{align}
and thus reproduces the loop quantum cosmology difference equation, see e.g. equation (3.8) in \cite{AshtekarRobustnessOfKey}. 
Up to this change of variables, \eqref{eq:QuantumHamiltonian} yields the same difference equation, with the loop quantum cosmology Hilbert space being embedded at a single point of the spatial slice where we have a non-trivial excitation of geometry and the scalar field. Due to implementing spatial diffeomorphisms, the information of where this point is gets erased in the reduced quantum states.

In \cite{AshtekarQuantumNatureOf}, the difference equation is slightly changed due to a different philosophy of the derivation. Taking the point of view that this equation should originate from the loop quantum gravity Hamiltonian constraint as defined by Thiemann \cite{ThiemannQSD1}, one of the factors of $\nu$ in \eqref{eq:HamiltonianReducedDeparabnu}, the right-most, should be replaced by $\left((\nu)^{-1}\right)^{-1} := \text{const} \cdot B(\nu)^{-1}$, where this inverse comes from regularising the term $P_\phi^2 / \sqrt{q}$ in the full Hamiltonian constraint. For large $\nu$, this expression converges to $\nu$, however some differences arise for small $\nu$. Similarly, the other $\nu$ in \eqref{eq:HamiltonianReducedDeparabnu}, the one in between the $\sin(\lambda b)$, is replaced by an expression $A(\nu)$, which again tends to $\nu$ for large $\nu$. This specific choice of difference equation in \cite{AshtekarQuantumNatureOf} is thus motivated by postulating that the regularisation of the Hamiltonian constraint, factor ordering, and choice of variables in \cite{ThiemannQSD1} is the correct one, and that in particular one should not rewrite the Hamiltonian classically in the deparametrised form $P_\phi \approx ...$. These assumptions can however be relaxed if one is sufficiently satisfied with the construction of the full theory Hamiltonian as given in appendix \ref{app:FullHamiltonian}, or if one wants to deparametrise already classically. 

The solutions to the difference equation and quantum Dirac observables are constructed along the same lines as in \cite{AshtekarQuantumNatureOf}. One first changes the scalar product to\footnote{This scalar product is ill-defined if $\nu = 0$. However, we can start with a quantum state that does not have support on $\nu = 4 \lambda \mathbb Z$. Due to the evolution equation \eqref{eq:bnuDifferenceEquation}, this is dynamically preserved.}
\be
	\braket{h^{\rho_\alpha, \rho_\phi}_{\text{diff}}}{h^{\rho_\alpha', \rho_\phi'}_{\text{diff}}}_\text{diff} = \frac{1}{|\rho_\alpha|} \delta_{\rho_\alpha,\rho'_\alpha} \delta_{\rho_\phi,\rho'_\phi} \text{,}
\ee
which makes $\tilde \Theta$, the analogue of $\Theta$ for the difference equation coming from \eqref{eq:QuantumHamiltonian}, a positive and self-adjoint operator. We can thus take its square root. The quantum states $e^{i (\phi-\phi_0) \sqrt{\tilde \Theta}} \ket{\rho_\alpha, 0}$ are in the kernel of the Hamiltonian constraint. Similarly, the operators $e^{i (\phi - \phi_0) \sqrt{\tilde \Theta}} \widehat{|\alpha|(\Sigma)} e^{-i (\phi - \phi_0) \sqrt{\tilde \Theta}}$ and $\widehat{ P_\phi(\Sigma)} = \sqrt{\tilde \Theta}$ commute with the quantum Hamiltonian constraint. Physical states are furthermore required to be symmetric under $\rho_\alpha \rightarrow - \rho_\alpha$.

\section{Comments} \label{sec:Comments}

\paragraph{Choice of symmetry reduced quantum states:} \mbox{}\\
A priori, there are several avenues along which a symmetry reduction in a full theory of quantum gravity could be constructed. In the case of a spatially flat FRW model, the only dynamical degrees of freedom (apart from matter) are the scale factor and its conjugate momentum, or similar variables as e.g. the $(b,v)$ variables from loop quantum cosmology. The first question that one might ask is how a symmetry reduced full theory state should look like. Should it be a very fine state with a suitable notion of homogeneity and isotropy, or should it be a maximally coarse state, which only carries the global gravitational and matter degrees of freedom? Clearly, given a very fine state, its total volume can be extracted, in principle along with an {\it effective} dynamical equation which it obeys. This effective dynamical equation does not need to agree with what one would obtain by applying the fundamental dynamics governing the fine state to a coarse state. While this first avenue is certainly the most satisfying and conceptually sound, it is also the hardest to follow. Recent progress along these lines can e.g. be found in \cite{GielenCosmologyFromGroup, AlesciLoopQuantumCosmology}. The second avenue, to use a very coarse state, is computationally simpler. A previous example of this approach is spinfoam cosmology \cite{BianchiTowardsSpinfoamCosmology}. The dynamics that one defines on the coarse state should be interpreted as an effective dynamics if one wants to use it to describe our {\it continuum} universe. As said before, it does not need to agree with the fundamental dynamics, which could act differently on a {\it single quantum} of volume of the size of the universe. Accordingly, it is not clear whether the full theory Hamiltonian constraint defined in appendix \ref{app:FullHamiltonian} has good semiclassical properties, since we used a regularisation which gives the correct effective dynamics, but didn't check physical viability beyond the context of the symmetry reduced single vertex states. In particular, a physically viable dynamics can depend on the coarse graining scale, see \cite{DittrichTheContinuumLimit} for further discussion.

\paragraph{Embedding into {\it the} full theory?:} \mbox{}\\
Given different quantisations based on different choices of variables, we have to be careful about the extent to which the question of deriving a certain theory as the symmetry reduced sector of the full theory can be meaningfully asked. In particular, a choice of full theory determines a certain preferred choice of point separating phase space functions which are quantised. Operators corresponding to other phase space functions can then subsequently be constructed by using classical relations. This process can however be highly ambiguous and result in operators with rather different properties. As an example \cite{BZI}, 
extracting the midi-superspace variables from SU$(2)$ holonomies and fluxes is a) generally rather complicated and b) not unique in the sense that no preferred way seems to exist. It is thus rather plausible that different effective dynamics will follow, with the details depending on the choice of embedding as well as the precise construction of the full theory dynamics.

Given these considerations, it seems clear that the question that can be meaningfully asked is the following: given a quantisation of a classically reduced theory, can we construct a symmetry reduction within a full quantum theory such that the two match for a suitable identification of classically and quantum reduced states? This question has been answered in the affirmative in this paper for the example of loop quantum cosmology in $(b,v)$ variables. For this, it was key in the derivation to pick a set of variables in the full theory such that the mini-superspace variables are included in the preferred set of fundamental full theory operators and that the other variables were chosen to commute with them.

\paragraph{Imposing homogeneity?:} \mbox{}\\
While the reduced quantum states considered in section \ref{sec:ReductionConstraintsOperators} are sufficient to extract the symmetry reduced quantum dynamics, homogeneity and isotropy are not imposed on them in a strict sense. In fact, imposing \eqref{eq:DiffPalpha} allows for arbitrary diffeomorphism-invariant correlations between the geometry and the scalar field if one goes beyond a single vertex. Part of the reduction thus seems to be achieved by the single vertex truncation.
To improve on this, one can consider a stronger condition than \eqref{eq:DiffPalpha}. One option is to impose $\alpha(R) = \alpha(R')$ for two regions $R$ and $R'$ of the same coordinate volume, and similar for $P_\phi$. This would naively lead to a reduced state of the form 
\be
	\ket{\exp \left(- i \rho_\alpha \sum_{x \in \Sigma} P_\alpha(x) \right) \exp \left( i \rho_\phi \sum_{x \in \Sigma} \phi(x) \right)}\text{,}
\ee	
however it seems hard to give a precise technical sense to this state as it is not a cylindrical function due to the dependence on an infinite number of point holonomies. Also, $\alpha(R)$ would need to be normalised by the number of points in $R$ to have a finite action. 

Another possibility is to impose that the two terms in \eqref{eq:DiffPalpha} vanish individually. In fact, imposing that $\alpha \partial_a P_\alpha$ vanishes is equivalent to $P_\alpha$ being constant on $\Sigma$ as long as $\alpha \neq 0$ (which is required in the classical phase space), and similar for $\phi$ (where no such condition on $P_\phi$ exists, apart from the intend of taking $\phi$ as a clock). For the reduced quantum states, this means that the rigging map construction for the diffeomorphism-invariant states has to be applied twice, resulting in the state
\be
	\ket{h_{\text{hom}}^{\rho_\alpha, \rho_\phi} } := \ket{ \left( \sum_{x \in \Sigma} e^{- i \rho_\alpha  P_\alpha(x)} \right) \left(\sum_{x' \in \Sigma}  e^{ i \rho_\phi  \phi(x')} \right)}\text{.} \label{eq:HomStates}
\ee	
The reduced operators of section \ref{eq:QuantisationDynamics} act on these states as on the single vertex states $\ket{h_{\text{diff}}^{\rho_\alpha, \rho_\phi}}$. However, states of the type \eqref{eq:HomStates} cannot encode any diffeomorphism invariant correlations between the matter and geometry sector, they are therefore even more reduced. Still, if going beyond a single vertex for the matter and / or geometry sector, these states contain additional information encoding how the total values of $\alpha(\Sigma)$ and $P_\phi(\Sigma)$ are composed from the individual vertices.

We chose not to use this type of states in the main text (as opposed to those of section \ref{sec:ReductionConstraintsOperators}), since it in general requires us to use the minisuperspace Hamiltonian, as opposed to a unit lapse full-theory Hamiltonian as sketched in appendix \ref{app:FullHamiltonian}. The problems involved become apparent when one considers a full-theory operator corresponding to $\int_\Sigma \left(P_\phi^2 / \sqrt{q}\right) d^3x$, which does not preserve the class of states \eqref{eq:HomStates}, whereas it preserves those of section \ref{sec:ReductionConstraintsOperators}.

\paragraph{Extension to Bianchi I:} \mbox{}\\
It is also interesting to consider an application of the variables derived here to Bianchi I models, as opposed to those in \cite{BIII}. In order to obtain the $\bar \mu$-scheme, we would need to change our quantisation prescription for the $\beta$ and $\gamma$-sectors slightly. The logic of arriving at the $\bar \mu$-scheme as described in section \ref{eq:QuantisationDynamics} tells us that we should again polymerise in such a way that deviations from the classical theory are expected as soon as the the matter energy density becomes large, meaning in the case of Bianchi I that $K_{xx} q^{xx}$, $K_{yy} q^{yy}$, and $K_{zz} q^{zz}$ remain much smaller than $1$. Since in the standard case one morally considers holonomies of $K_{x i} = K_{xx} e^x_i$, one needs to choose $\lambda$ to be the inverse size of the universe in $x$-direction, and similar for $y$ and $z$. Next to the arguments in \cite{AshtekarLoopQuantumCosmologyBianchi}, this becomes very clear when considering the full-theory embedding \cite{BIII}. Returning to our variables, we would thus construct holonomies form $\beta$ and $\gamma$, integrated over regions $R$. Invariance under the diffeomorphisms coming from the reduction constraints then tells us that these regions need to be all of $\Sigma$. In order for the approximations $\beta(\Sigma) \approx \sin \left( \lambda_\beta \beta(\Sigma) \right) / \lambda_\beta$ to be valid away from Planck energy density, we need to set $\lambda_\beta \sim 1/ \rho_\alpha$, and similar for $\gamma$. To see why, note that $\beta$ and $\gamma$ are linear combinations of $K_{xx} q^{xx}$, $K_{yy} q^{yy}$, and $K_{zz} q^{zz}$ multiplied by $\alpha$. For further research, it would be interesting to compare the so-defined dynamics to the one given in \cite{AshtekarLoopQuantumCosmologyBianchi} and check for possible simplifications.

\section{Conclusion} \label{sec:Conclusion}

The main point of this paper was to show that loop quantum cosmology formulated in $(b,v)$ variables can be embedded into a full theory setting. For this, the full theory had to be constructed using different variables than in standard loop quantum gravity, however following the same quantisation strategy. The goal of extracting the loop quantum cosmology dynamics from the action of the full theory Hamiltonian on a set of symmetry reduced full theory quantum states was achieved. For this, the choice of quantisation variables was essential. Comments on the conceptual setup we are working in, in particular the currently unknown relation of effective and fundamental dynamics, have been given in the previous section. In particular, it is so far unclear whether the full theory dynamics we defined with the goal of reproducing the classically reduced quantisation is viable at a fundamental level, e.g. for finer states. We leave this important question for further research.

\section*{Acknowledgements}
This work was supported  by the Polish National Science Centre grant No. 2012/05/E/ST2/03308. We thank J\k{e}drzej \'Swie\.zewski for proofreading this manuscript.

\begin{appendix}

\section{Full theory Hamiltonian constraint} \label{app:FullHamiltonian}

In this appendix, we will describe how a quantisation of the full theory Hamiltonian constraint can be constructed such that it reproduces the mini-superspace Hamiltonian constraint when acting on a single-vertex type state as used in section \ref{sec:ReductionQuantisation}. Technically, our Hilbert space will be somewhat different in that we will use a fixed lattice for reasons discussed below. On a conceptual and qualitative level however, the case of a single lattice site will correspond to our single-vertex type reduced states used in the extraction of the dynamics. Our aim in this appendix will not be to construct a Hamiltonian constraint operator which is satisfactory from a full theory perspective, i.e. anomaly free and allowing for a large enough kernel with good semiclassical properties. Our only aim will be to show that, following previous quantisation techniques developed within loop quantum gravity, one can arrive at a plausible operator whose action reduces to the loop quantum cosmology difference equation when acting on a maximally coarse state. 

The definitions \eqref{eq:NewVars1}, \eqref{eq:NewVars2}, and \eqref{eq:NewVars3} suggest a natural smearing of our variables. $\alpha, \beta$, and $\gamma$, being densities, will be smeared over three-dimensional regions, while $P_\alpha$, $P_\beta$, and $P_\gamma$ will be exponentiated to point holonomies without further smearing, as already done in section \ref{sec:QuantumKinematics}. In order to quantise the full Hamiltonian constraint, one needs to reconstruct the ADM canonical variables by inverting \eqref{eq:NewVars1}, \eqref{eq:NewVars2}, and \eqref{eq:NewVars3}, which gives 
\begin{align}
	q_{xx} &= \left(\alpha^2 e^{-2 P_\beta -2 P_\gamma} \right)^{\frac13}, ~ &P^{xx}q_{xx} = \frac 12 P_\alpha \alpha + \frac 13 \beta + \frac 13 \gamma \\
	q_{yy} &= \left(\alpha^2 e^{+4 P_\beta -2 P_\gamma} \right)^{\frac13}, ~ &P^{yy}q_{yy} = \frac 12 P_\alpha \alpha - \frac 23 \beta + \frac 13 \gamma \\
	q_{zz} &= \left(\alpha^2 e^{-2 P_\beta +4 P_\gamma} \right)^{\frac13}, ~ &P^{zz}q_{zz} = \frac 12 P_\alpha \alpha + \frac 13 \beta - \frac 23 \gamma \text{,}
\end{align}
and thereby express the Hamiltonian constraint in our new variables. 

A first choice to be made is whether one wants to artificially introduce a fixed lattice, i.e. a fixed underlying graph. The choice of a fixed lattice, or superpositions thereof, are helpful in promoting spatial derivatives to operators, since one can evaluate fields at neighbouring lattice sites and take their difference. A similar picture also appears within loop quantum gravity automatically without the introduction of a lattice, however with somewhat different properties of the action of the Hamiltonian constraint \cite{ThiemannQSD1, ThiemannQSD5}. So far, we were not able to construct a  full theory constraint operator without the introduction of a fixed lattice which would at the same time reproduce the loop quantum cosmology difference equation {\it and} be sufficiently plausible on generic, non-reduced states. Therefore, we will only outline a construction based on a fixed lattice approximation in this paper. 

While in Thiemann's original construction of the Hamiltonian constraint operator one approximates the spatial integral in the Hamiltonian constraint by a Riemann sum, regulates the constraint at finite discretisation, promotes it to an operator, and subsequently removes this regulator by infinitely refining the discretisation, one cannot follow this last step when introducing a fixed lattice. Instead, the fundamental lattice, or, in other words, the underlying graph, sets the ``scale'' until which the discretisation can be refined. Such an approach has for example been taken in the algebraic quantum gravity framework \cite{GieselAQG1}. 

As a lattice, or fixed underlying graph, we will employ a regular cubulation of the three-torus with $N_x N_y N_z$ lattice sites, or graph vertices. At these vertices, we have point holonomies of $P_\alpha$, $P_\beta$, $P_\gamma$, and $\phi$. $\alpha$, $\beta$, $\gamma$, and $P_\phi$ are smeared over the cubes dual to the vertices. Using this smearing, one can arrive at an expression for the Hamiltonian constraint regularised on the prescribed lattice. Derivatives are regularised as finite differences for nearest neighbours. Inverse powers of $\alpha$ are regulated via Thiemann's Poisson bracket trick \cite{ThiemannQSD1}. $P_\beta$ and $P_\gamma$ have to be regulated as point holonomies as, e.g.,  $\sin{ (\lambda P_\beta)} / \lambda$. Suitable inverse powers of other components of the metric can then be regulated as products of (inverse) $\alpha$ and, e.g., $\exp \left(\pm \sin{ (\lambda P_\beta)} / \lambda \right)$.

We will not spell out the details of this regularisation in this paper, as it is not of importance for the main message. We can however already infer the central conclusion of this appendix. For this, we consider a lattice which has just a single lattice site, so that this lattice site is its own neighbour in all three directions. Such a state corresponds to the single-vertex reduced state from section \ref{sec:ReductionQuantisation}. As in section \ref{sec:ReductionQuantisation}, we choose a quantum state such that the action of $\beta$ and $\gamma$ at this vertex vanishes. All derivatives also vanish, since their finite difference regularisation results in terms of the form $1-1=0$ due to the single vertex being its own neighbour. For the $P_{ab}^{\text{tf}} P^{ab}_{\text{tf}} / \alpha$-term, we choose an ordering such that $\beta$ and $\gamma$ are ordered to the right, so that the action of this term vanishes. 

It remains to deal with a term proportional to the spatial diffeomorphism constraint coming from stabilising the diagonal metric gauge. This term is problematic, because in order to know it explicitly, one has to invert a coupled partial differential equation. In the quantum reduction, this was avoided, since all corresponding spatial diffeomorphisms acting on the reduced phase space were constrained to vanish by the reduction constraints and the set of all diffeomorphisms could be implemented by standard LQG techniques. In the full theory however, it remains. Still, since all terms in the spatial diffeomorphism constraint are proportional to derivatives, we can again argue that, as above, a suitable regularisation of the constraint would vanish on our maximally coarse states, independently of how exactly the partial differential equation is inverted. This is also true for the diffeomorphisms acting as constant shifts, which preserve our gauge fixing.

We are thus only left with the two terms already treated in section \ref{eq:QuantisationDynamics}. Here, we have again some freedom in whether we want to deparametrise before or after quantisation, i.e. quantise the original Hamiltonian, or one rewritten in the from $P_\phi - H_{\text{true}} = 0$. This again only affects some details of the difference equation, e.g. the previously discussed difference of having $\nu$ or $\text{const} \cdot B(\nu)^{-1}$ appearing. 

We conclude that it is possible to prescribe a plausible quantisation of the full theory Hamiltonian constraint on a fixed underlying graph. If one chooses this graph to be maximally coarse, i.e. to consist only of a single vertex, and sets the $\beta$ and $\gamma$ parts of the quantum state to zero, one recovers the loop quantum cosmology difference equation.

\section{Spherical Symmetry} \label{app:SphericalSymmetry}

\subsection{Classical preparations}

A construction very similar to the one in the main part of this paper can also be repeated for spherical symmetry. We will be less explicit with the details here, because the extension to spherical symmetry is straight forward building on earlier work \cite{BZI, BLSI}. 

As outlined in \cite{BZI}, we implement the radial gauge for the spatial diffeomorphism constraint, retaining radial diffeomorphisms whose shift vector is independent of the angular variables. The three coordinates on the spatial slice are split into a radial coordinate $r$ and two angular coordinates collectively denoted by $\theta$, whose corresponding tensor indices are denoted by $A,B$. After the gauge fixing $q_{rA} = 0$ and $q_{rr}(r,\theta) = q_{rr}(r)$, the free variables and their non-vanishing Poisson brackets are $\{q_{rr}(r, \theta), P^{rr}(r',\theta')\} = \delta(r,r')\delta^{(2)}(\theta,\theta')$ and $\left\{ q_{AB}(r,\theta),  P^{CD}(r',\theta') \right\} = \delta_{(A}^C \delta_{B)}^D \delta(r,r') \delta^{(2)}(\theta,\theta')$.

For spherical symmetry, the analogue of implementing $P^{a \neq b} = 0$ is to implement $P^{rA} = 0$. After solving $C_A$ for $P^{rA}$, we find that \cite{BZI, BLSI}
\be
	\int dr d^2 \theta \, P^{rA} c_A = 0~  \forall ~ c_A(x) ~~ \Leftrightarrow ~~ \int dr d^2 \theta \, \left( P^{rr} \mathcal L_N q_{rr} + P^{AB} \mathcal L_N q_{AB} \right) = 0 ~ \forall ~ N^A(x) \text{,} \label{eq:PrA}
\ee
where $\mathcal L_N$ denotes the Lie derivative w.r.t. the vector field $N^A$ (with only angular indices, i.e. $N^r = 0$), $q_{AB}$ behaves as a 2-metric under the Lie derivative, and $q_{rr}$ as a scalar. 
In other words, $P^{rA}$ acts as the generator of spatial diffeomorphisms in angular directions on the variables in the reduced phase space. 

Next, we need to define suitable variables, analogous to those for the FRLW model:
\begin{align}
		\Lambda & = \sqrt{q_{rr}} ~ & P_\Lambda &:= 2 \sqrt{q_{rr}} P^{rr}- \frac{P^{\theta \theta} q_{\theta \theta}+P^{\phi \phi} q_{\phi \phi}}{\Lambda} \\
		\alpha &:= \sqrt{\Lambda^2  q_{\theta \theta} q_{\phi \phi}} ~ &P_\alpha &:= \frac{P^{\theta \theta} q_{\theta \theta}}{\sqrt{\Lambda^2  q_{\theta \theta} q_{\phi \phi}}}+\frac{P^{\phi \phi} q_{ \phi \phi}}{\sqrt{\Lambda^2  q_{\theta \theta} q_{\phi \phi}}} \\
		\beta &:=  \frac{1}{2} \left(P^{\theta \theta} q_{\theta \theta} -P^{\phi \phi} q_{\phi \phi} \right) ~~ &P_\beta &:= \log \frac{q_{\phi \phi}}{q_{\theta \theta}} \label{eq:BetaDefintion}\\
		\gamma &:= P^{\theta \phi} & P_\gamma &:=- q_{\theta \phi}
\end{align}
and compute the new non-vanishing Poisson brackets
\begin{align}
	&\left\{ \Lambda (x), P_\Lambda(y) \right\} = \delta^{(3)}(x,y) ~~ \text{and} \\
	&\left\{ \alpha (x), P_\alpha(y) \right\} = \delta^{(3)}(x,y), ~~~ \left\{ \beta (x), P_\beta(y) \right\} = \delta^{(3)}(x,y),  ~~~ \left\{ \gamma (x), P_\gamma(y) \right\} = \delta^{(3)}(x,y) \text{.}
\end{align}

The spherically symmetric spatial line element in standard spherical coordinates reads
\be
	ds^2 = \Lambda(r)^2 dr^2 + R(r)^2 (d \chi^2 + \sin^2 \chi d \phi^2) \text{.}
\ee
The canonically conjugate variables are $P_\Lambda(r)$ and $P_R(r)$. 
By symmetry considerations, the momentum $P^{ab}$ has to be proportional to $\sqrt{q} q^{ab}$ up to a radial dependence. It follows directly that in the sector of the full phase space corresponding to the midi-superspace, we have 
\be
	\beta = 0, ~~~ P_\beta - \log \sin \chi = 0, ~~~ \gamma = 0, ~~~ P_\gamma = 0. \label{eq:FirstReductionConstraints}
\ee
We thus add \eqref{eq:FirstReductionConstraints} to our list of reduction constraints. We can now form linear combinations with \eqref{eq:PrA} and instead of \eqref{eq:PrA} use
\be
	\tilde C_A[N^A] := \int  dr d^2 \theta  \, \left( P_\Lambda \mathcal L_N \Lambda +   P_\alpha \mathcal L_N \alpha \right) \label{eq:CTildeA}
\ee
as a reduction constraint. 
Similarly\footnote{In \eqref{eq:CrFirst}, we again dropped terms containing $\beta$ and $\gamma$ for simplicity. This is conceptually somewhat different from deriving $\tilde C_A$, because $C_r$ is not a reduction constraint. However, even without dropping those terms, the action of \eqref{eq:CrFirst} is the same on cylindrical functions not depending on $P_\beta$ and $P_\gamma$, so that the derivation works as before even if we keep those terms. \label{ftn:SuperposeConstraints}}, we also rewrite the radial part of the spatial diffeomorphisms using \eqref{eq:FirstReductionConstraints} as
\be
	C_r[N^r] = \int dr d^2 \theta  \, \left( P_\Lambda \mathcal L_N \Lambda +   P_\alpha \mathcal L_N \alpha \right) \text{,} \label{eq:CrFirst}
\ee	
where $\Lambda$ and $\alpha$ behave as radial densities and $N^r = N^r(r)$.

\subsection{Quantum theory}

The construction of the quantum theory proceeds largely analogously to the case of cosmology, the main difference being the smearing directions. For the $\alpha$ and $\beta$ sectors, the kinematical quantisation is the same as above. 
In the $\Lambda$ sector, we proceed differently because we cannot turn $\Lambda$ into a density with respect to both radial and angular directions as it would coincide, up to normalisation, with $\alpha$ in this case. We thus smear $\Lambda$ along radial lines and $P_\Lambda$ over surfaces embedded into the spheres $S^2_r$ of constant geodesic distance form the centre. Next to the dependence on $P_\alpha$ via point holonomies, a cylindrical function will thus also depend on a finite number of arguments of the form $\exp\left(-i \rho_\Lambda \int_{A_r} P_\Lambda \right)$, where $A_r$ is a surface embedded into an $S^2_r$ and $\rho_\Lambda \in \mathbb R$. 
Implementing the angular diffeomorphisms resulting from $P^{rA}=0$ can be most conveniently\footnote{More general solutions are possible, such as a diffeomorphism equivalence class as in the $\alpha$ sector. However, for the purpose of extracting the midisuperspace dynamics, the choice $A_r = S^2_r$ is sufficient.} done by choosing $A_r$ to coincide with $S^2_r$. 
In the $\gamma$ sector, the index structure suggests to smear $P_\gamma$ over a radial line and $\gamma$ over a surface $A_r$ as above. 
Cylindrical functions then additionally depend on a finite number of arguments of the form $\exp \left(-i \rho_\gamma \int_{r_1}^{r_2} dr P_\gamma \right)$.

Out of the reduction constraints \eqref{eq:FirstReductionConstraints}, we choose to implement $\beta=0$ and $\gamma=0$ as first class constraints. They enforce that the reduced cylindrical functions do not depend on $P_\beta$ or $P_\gamma$. At this point, a cylindrical function obeying the reduction constraints thus depends on a finite number of arguments of the form $\exp({- i \rho_\alpha P_\alpha})(r)$ and $\exp \left(-  i \rho_\Lambda \int_{S^2_r} P_\Lambda \right)$. The $r$-dependency in the point holonomy of $P_\alpha$ indicates that the angular dependence is averaged out by a diffeomorphism average over $S^2_r$ as in the main part of this paper. 

A convenient basis in the set of symmetry reduced cylindrical functions is given by functions of the type 
\be
	\ket{\rho_\alpha^{r_1}, \ldots, \rho_\alpha^{r_n}; \rho_\Lambda^{r_1}, \ldots, \rho_\Lambda^{r_n}} := e^{-i \rho_\alpha^{r_1} P_{\alpha}} (r_1) \ldots e^{-i \rho_\alpha^{r_n} P_{\alpha}} (r_n) \cdot e^{-i \rho_\Lambda^{r_1} \int_{S^2_{r_1}} P_{\Lambda}}  \ldots e^{-i \rho_\Lambda^{r_n} \int_{S^2_{r_n}} P_{\Lambda}} 	 \text{.}
\ee
The action of point holonomies of $P_\alpha$ and $\int_{S^2_{r}} P_{\Lambda}$ thereon is by multiplication as above. $\Lambda(e) := \int_{r_a}^{r_b} dr \Lambda(r,\theta) $ and $\alpha(R) := \int_{r_a}^{r_b} dr  \int_{S^2_r} d^2 \theta \, \alpha$ act as
\begin{align}
	\widehat {\Lambda(e)} \ket{\rho_\alpha^{r_1}, \ldots, \rho_\alpha^{r_n}; \rho_\Lambda^{r_1}, \ldots, \rho_\Lambda^{r_n}} &= \sum_{i=1}^n \, s([r_a, r_b], r_i) ~ \rho_\Lambda^{r_i}  \ket{\rho_\alpha^{r_1}, \ldots, \rho_\alpha^{r_n}; \rho_\Lambda^{r_1}, \ldots, \rho_\Lambda^{r_n}}\\
		\widehat {\alpha(R)} \ket{\rho_\alpha^{r_1}, \ldots, \rho_\alpha^{r_n}; \rho_\Lambda^{r_1}, \ldots, \rho_\Lambda^{r_n}} &= \sum_{i=1}^n \,  s([r_a, r_b], r_i) ~ \rho_\alpha^{r_i}\, \ket{\rho_\alpha^{r_1}, \ldots, \rho_\alpha^{r_n}; \rho_\Lambda^{r_1}, \ldots, \rho_\Lambda^{r_n}}
\end{align}
where
\be
	 s([r_a,r_b], r) := \sign (r_{2} - r_{1}) ~ \times ~ \begin{cases} 
												1 &\mbox{if } r \in (r_a, r_b) \\ 
												1/2 & \mbox{if } r = r_a ~~ \text{or}~~ r = r_b \\  
												0 & \mbox{otherwise.}					
											\end{cases} 
\ee

On these symmetry reduced cylindrical functions, we now implement the radial spatial diffeomorphism constraint, which effectively turns the $r$-labels in the cylindrical functions into an ordered set of radial lattice points. 
A symmetry reduced cylindrical function with implemented radial spatial diffeomorphisms and non-trivial support at $n$ radial lattice sites (whose order, but not their location, is their only intrinsic property) is thus labelled by a set of quantum numbers $\rho_\alpha^1, \ldots, \rho_\alpha^n$ and $\rho_\Lambda^1, \ldots, \rho_\Lambda^n$. 

We can now relate the midisuperspace degrees of freedom $\Lambda(r), P_\Lambda(r), R(r)$, and $P_R(r)$ to the operators which commute with the reduction constraints. First, we are going to change variables in the midisuperspace formulation to $\tilde P_\Lambda := P_\Lambda - \frac{P_R R}{2 \Lambda}$, $v := \Lambda R^2$, and $P_v := \frac{P_R}{2 \Lambda R}$, since these variables are the direct analogues of our full theory variables. Their non-vanishing Poisson brackets are $\{ \Lambda(r), \tilde P_\Lambda(r') \} = \delta(r,r')$ and $\{ v(r),  P_v(r') \} = \delta(r,r')$. 
The following relations hold between the full phase space and the midisuperspace variables:
\begin{align}
	\int_{r_1}^{r_2} dr \Lambda(r) &= \int_{r_1}^{r_2} dr \Lambda(r,\theta) 	\label{eq:RelLambdaNuFR1} \\
	\int_{r_1}^{r_2} dr \, v(r) &= \frac{1}{4\pi}\int_{r_1}^{r_2} dr \int_{S^2_r} d^2 \theta \, \alpha \label{eq:RelLambdaNuFR2}\\
	\int_{r_1}^{r_2} dr  P_v(r) v(r) &= \int_{r_1}^{r_2}  dr \int_{S^2_r} d^2 \theta \,  P_\alpha \alpha \label{eq:RelLambdaNuFR3} \\
	\int_{r_1}^{r_2} dr  \tilde P_\Lambda(r) \Lambda(r) &= \int_{r_1}^{r_2}  dr \int_{S^2_r} d^2 \theta \, P_\Lambda(r,\theta) \Lambda(r,\theta)  \label{eq:RelLambdaNuFR4}
\end{align}

We are now in a position to express the midisuperspace Hamiltonian constraint
\begin{align}
	H &= \frac{1}{16 \pi v} \left( (\tilde P_\Lambda \Lambda )^2 - 3(P_v v )^2 - 2 P_\Lambda \Lambda P_v v \right) \nonumber\\
	&~~~~ + 8\pi \left( -\frac{1}{2} \Lambda + \frac{11}{8} \frac{(\Lambda')^2 v}{\Lambda^4} - \frac{5}{4} \frac{\Lambda' v'}{\Lambda^3 } - \frac{1}{2} \frac{\Lambda'' v}{\Lambda^3 } - \frac{1}{8} \frac{(v')^2}{\Lambda^2 v } + \frac{1}{2} \frac{ v''}{\Lambda^2 }  \right) \label{eq:HamLambdaNu}
\end{align}
using operators acting on the symmetry reduced cylindrical functions. The regularisation proceeds along standard lines. We approximate the radial integral in the Hamiltonian constraint $H[N] = \int_0^\infty dr N H$ by a Riemann sum, later taking the limit of an infinitely fine discretisation. 
We then approximate, e.g., 
\be
	\int_{r-\epsilon}^{r+\epsilon} dr' \frac{(\tilde P_\Lambda \Lambda )^2}{v} \approx \frac{\left(\int_{r-\epsilon}^{r+\epsilon} dr' \, \tilde P_\Lambda \Lambda \right)^2}{\int_{r-\epsilon}^{r+\epsilon} dr' \, v} \text{.}
\ee
The full theory analogue of $\int_{r-\epsilon}^{r+\epsilon} dr' \, \tilde P_\Lambda \Lambda$ is given by $\int_{r-\epsilon}^{r+\epsilon}  dr \int_{S^2_r} d^2 \theta \, P_\Lambda(r,\theta) \Lambda(r,\theta)$, which furthermore needs to be polymerised as, e.g.,
\be 
	\int_{r-\epsilon}^{r+\epsilon} dr' \, \tilde P_\Lambda \Lambda \approx 	 \frac{1}{\lambda} \sin \left( \lambda \int_{S^2_r} d^2 \theta \, P_\Lambda(r,\theta) \right) \int_{r-\epsilon}^{r+\epsilon}  dr' \Lambda(r') \label{eq:PreOpPLL}
\ee
\eqref{eq:PreOpPLL} can now be promoted to an operator $\widehat  { P_\Lambda \Lambda (e_r^\epsilon)}$, which acts as
\begin{align}
	&\widehat  { P_\Lambda \Lambda (e_r^\epsilon)} \ket{\rho_\alpha^{r_1}, \ldots, \rho_\alpha^{r_n}; \rho_\Lambda^{r_1}, \ldots,  \rho_\Lambda^{r_n}} \!\!\!\!\!\!\!\!\!\!\!\!\!\!\!\!\!\!\!&& \\ \nonumber
	= &\frac{1}{2 i \lambda} \sum_{i=1}^n \, s([r-\epsilon, r+\epsilon], r_i) ~ \rho_\Lambda^{r_i}  \Biggl(&&+ \ket{ \rho_\alpha^{r_1}, \ldots, \rho_\alpha^{r_n}; \rho_\Lambda^{r_1}, \ldots, \rho_\Lambda^{r_{i-1}}, \rho_\Lambda^{{r_i}-\lambda}, \rho_\Lambda^{r_{i+1}}, \ldots, \rho_\Lambda^{r_n}} \\ \nonumber
	&&&- \ket{ \rho_\alpha^{r_1}, \ldots, \rho_\alpha^{r_n}; \rho_\Lambda^{r_1}, \ldots, \rho_\Lambda^{r_{i-1}}, \rho_\Lambda^{{r_i}+\lambda}, \rho_\Lambda^{r_{i+1}}, \ldots,\rho_\Lambda^{r_n}} \Biggr) \text{.}
\end{align}
Similarly, $\int_{r-\epsilon}^{r+\epsilon} dr' \, \tilde P_v v = \int_{r-\epsilon}^{r+\epsilon} dr' \int_{S^2_{r'}} d^2 \theta \, P_\alpha(r,\theta) \alpha(r,\theta)$ is polymerised as \newline $\int_{r-\epsilon}^{r+\epsilon} dr'   \int_{S^2_{r'}} d^2 \theta \,\frac{1}{\lambda} \sin\left(\lambda P_\alpha(r',\theta) \right) \alpha(r',\theta)$ and the corresponding operator $\widehat{P_\alpha \alpha(e^\epsilon_r)}$ acts as 
\begin{align}
	&\widehat  { P_\alpha \alpha (e_r^\epsilon)} \ket{\rho_\alpha^{r_1}, \ldots, \rho_\alpha^{r_n}; \rho_\Lambda^{r_1}, \ldots,  \rho_\Lambda^{r_n}} \!\!\!\!\!\!\!\!\!\!\!\!\!\!\!\!\!\!\!&& \\ \nonumber
	= &\frac{1}{2 i \lambda} \sum_{i=1}^n \, s([r-\epsilon, r+\epsilon], r_i) ~ \rho_\alpha^{r_i}  \Biggl(&&+ \ket{\rho_\alpha^{r_1}, \ldots, \rho_\alpha^{r_{i-1}}, \rho_\alpha^{{r_i}-\lambda}, \rho_\alpha^{r_{i+1}}, \ldots, \rho_\alpha^{r_n}; \rho_\Lambda^{r_1}, \ldots, \rho_\Lambda^{r_n}} \\  \nonumber
	&&&- \ket{\rho_\alpha^{r_1}, \ldots, \rho_\alpha^{r_{i-1}}, \rho_\alpha^{{r_i}+\lambda}, \rho_\alpha^{r_{i+1}}, \ldots, \rho_\alpha^{r_n}; \rho_\Lambda^{r_1}, \ldots, \rho_\Lambda^{r_n}} \Biggr) \text{.}
\end{align}
The factor $1/v$ in the Hamiltonian can be regulated again via Poisson bracket tricks. In order to preserve the single vertex structure of the $\alpha$ sector (at every non-trivial $r_i$) of our symmetry reduced quantum states, we rewrite $ \frac{1}{v}$ as  $\frac{1}{v^{1+\delta}} v^\delta$ and quantise this expression in the suggested ordering. Since $\widehat {v^\delta}$ vanishes if it acts at a point with no non-trivial $e^{i P_\alpha}$ dependence, the resulting operator for $\frac{1}{v^{1+\delta}} v^\delta$ does so, too. As in the main text, we can also introduce absolute values around $\alpha$ and $\Lambda$ in order to ensure their positivity. 

The quantisation of the second line of \eqref{eq:HamLambdaNu} is again straight forward with the methods discussed so far. We only have to pay attention to order a factor of $v$ to the right so that the resulting expression preserves our symmetry reduced states. 

The action of the resulting Hamiltonian constraint operator then agrees with the action that one would have derived form a midisuperspace quantisation. Since our strategy was to quantise the midisuperspace Hamiltonian as an operator on the symmetry reduced Hilbert space of the full theory, this is not very surprising because the algebra of elementary operators of the midisuperspace quantisation is reproduced when expressing them via the relation \eqref{eq:RelLambdaNuFR1}, \eqref{eq:RelLambdaNuFR2}, \eqref{eq:RelLambdaNuFR3}, and \eqref{eq:RelLambdaNuFR4} with full theory operators. Also, their action on the quantum states agrees when one straight forwardly maps quantum states form the midi-superspace quantisation to symmetry reduced states in the full theory.

We can now again repeat the discussion of appendix \ref{app:FullHamiltonian} and construct a full theory Hamiltonian, based on a fixed underlying graph, so that when restricting the number of lattice sites to a single vertex per non-trivial $S^2_r$, we again obtain the midisuperspace Hamiltonian as a full theory operator. While it seems plausible that this can be done, we have not looked into this in detail so far.

Also, one can ask now to which extent one has implemented a $\bar \mu$ scheme for spherical symmetry. The $\alpha$ sector of our variables behaves exactly like in the context of cosmology in the main part of the paper, that is $P_\alpha$ is scalar. Our construction there is thus the direct analogue of the $\bar \mu$ scheme in $(b,v)$ variables for spherical symmetry. In the $\Lambda$ sector however, $P_\Lambda$ has to be smeared over the $S^2_r$. We could not construct $P_\Lambda$ as a scalar because then it would have to be conjugate to a multiple of $\alpha$. Therefore, a naive\footnote{In LQC, a possible motivation for the $\bar \mu$ scheme is that the polymerisation scale $\lambda^{-1}$ cuts off the integrated extrinsic curvature $\int K_a ds^a$, see the derivation in \cite{BIII}. However, $\int K_a ds^a \propto \sqrt{\rho_\phi} \times \text{distance}$ for a spatially flat FRW model, where $\rho_\phi$ is the matter energy density. Therefore, in order to get corrections only when the energy density approaches the Planck density, one needs to use a polymerisation scale according to the $\bar \mu$ scheme, i.e. $\lambda^{-1} \propto \text{distance}$.} extension of the $\bar \mu$ scheme as formulated for $(c,p)$ variables \cite{AshtekarQuantumNatureOf} would be to set the polymerisation scale $\lambda = \lambda(r) \propto   R(r)^{-2}$ in the $\Lambda$ sector, while keeping $\lambda$ fixed in the $\alpha$ sector. 
Whether the so defined dynamics has good semiclassical properties is so far unclear, since fully dynamical calculations based on spherically symmetric loop quantum gravity are scarce at this point, see e.g. \cite{BojowaldLemaitreTolmanBondi, BojowaldNonmarginalLemaitreTolman}.


\end{appendix}

\end{document}